\documentclass[vecphys]{svmult}
\usepackage{amsmath}
\usepackage{amsfonts}
\usepackage{amssymb}

\usepackage{makeidx}         
\usepackage{graphicx}        
\usepackage{multicol}        
\usepackage[bottom]{footmisc}

\makeindex             

\begin{document}

\title*{Chaos in Galaxies}
\author{Daniel Pfenniger}
\institute{Geneva Observatory, University of Geneva, 1290 Sauverny, Switzerland}
%

\maketitle

\abstract{After general considerations about limits of theories and
  models, where small changes may imply large effects, we discuss three cases in galactic astrophysics illustrating how galactic dynamics models may become insufficient when previously neglected effects are taken into account: 
\begin{enumerate}
 \item Like in 3D hydrodynamics, the non-linearity of the Poisson-Boltzmann system may imply dissipation through the growth of discontinuous solutions. 
 \item The relationship between the microscopic exponential sensitivity of N-body systems and the stability of mean field galaxy models.
 \item The role of quantum physics in the dynamics of structure formation, considering that cosmological neutrinos are massive and semi-degenerate fermions.
\end{enumerate}

}

\section{Introduction}

\subsection{Chaos in science}
Chaos plays a key role in many sciences, but particularly in galactic astrophysics where it appears under different aspects in various galaxy models. Before discussing chaos in galaxies, some space will be spent  discussing chaos in science. 

The notion of chaos has deeply modified the scientist view of Nature, but also of the scientific process itself.  But what makes chaos so special?   In sciences chaos expresses commonly two important properties of \textit{dynamical models} of natural phenomena: 
\begin{enumerate}
 \item \textit{The sensitivity to initial conditions}.  Initially close solutions of a dynamical model separate in average at least exponentially with time, $\langle \lVert\Delta x (t)\rVert\rangle > \exp (t/\tau)$, such that after a few characteristic Liapunov time $\tau$ the small scale effects neglected either in the model, or in the numerical model of the mathematical model, become macroscopic.  This restricts a deterministic use of chaotic models over a model dependent finite time, and means that in chaotic systems  Laplace's determinism is only applicable approximately over a limited time.
 \item \textit{The rapid mixing of the model solutions}. An exponential divergence of nearby solutions is not sufficient to produce chaos in the technical sense.  Indeed, if a uniformly diverging flow preserves the neighbourhood of solutions,  a smooth transformation of coordinates allows to remove the formal divergence back: the regularized model is then no longer diverging.  Therefore the divergence effect may be model dependent. To be truly chaotic, a system should remain chaotic even after smooth transformations of coordinates.  It should \textit{mix} the solutions, such that after a few Liapunov times not only initially close solutions become widely separated by the exponential divergence, but some initially distant solutions come close together at some time.  With time, most of solutions become at least once arbitrarily close to each other, which expresses the mixing property of chaotic systems.  Mixing occurs frequently in systems with bounded phase space, while in systems with unbounded phase space escaping solutions, i.e., solutions going to infinity, do not return often in almost all phase space regions. 
\end{enumerate}

Chaos is therefore important because it touches to an essential aspect of the scientific activity: to represent faithfully natural phenomena with formal models.  At the heart of the scientific process is the determination of the domain of applicability and the limits of models and theories.  

Actually we can enlarge the notion of sensitivity of solutions to initial conditions to the sensitivity of solutions to functionally close dynamical systems.  The former case is a particular case of the latter one when the initial displacement has been produced by an initial impulse perturbation.  Therefore in chaotic systems not only possible perturbations of the initial conditions lead to unpredictability, but also close but distinct functional approximations adopted in the model may lead to very different solutions after a finite time.  

Therefore when considering chaos in natural systems, it is not only important to discuss how the ignorance of the real initial conditions influences the model predictions, but also how the functional ``distance'' from the real system to the modelled one participates to unpredictability.
This point is actually central to physical modelling, because a good model should be robust to perturbations of the model functional form, since a model is always an approximation of the real system. 

\subsection{Epistemological digression, theories vs. models}
Until recent times theories and models could be viewed as fundamentally different: physical theories, like quantum physics or general relativity, have been thought to be more fundamental than models, because applicable to many more different cases.  For a long time, since about the rise of classical mechanics in the 17--18$^{\mathrm{th}}$ centuries, theories were even taken as absolutely exact.  Isaac Newton's ``laws'' were widely considered as absolute at the same level as a God given, revealed truth.  

In the meanwhile, classical mechanics was found to be only an approximation of Nature, and had to be corrected by relativistic and quantum effects.  Further, the new physical theories also could not claim to be exact, because quantum physics and general relativity have remained incompatible since then, the theory of quantum gravity is still a work in progress.  The so-called fundamental physical theories, like the elementary particle Standard Model, are today known to be incomplete, and must all be considered as approximations of Nature.

Models, like a galaxy or star model, concern often more specific phenomena than theories. Models are simplified formal approximations of particular phenomena, while theories aimed in the past to be exact isomorphisms with reality, but aren't.  In models, scientists deliberately simplify reality in order to keep essential features and discard inessential ones.  Doing so allows to describe in a formal way, with usual mathematics but more and more with computer programs, the gist of the phenomena.  For example, a planet or a star may be approximated first by a perfect sphere, which allows to concentrate the discussion on its most essential features that we want to understand, and leave out the inessential aspherical features, like rotation induced flattening, or mountains.  This simplification or reality has the important virtue to reduce the complexity of phenomena to a level compatible with the finite capacity of human brains to grasp complex systems.  In the end, the subjective feeling of understanding a phenomenon comes from a simplified, i.e., inexact, description that captures essential features and discards inessential ones.  This essential-inessential separation is often not unique, contains subjective assumptions, and of course is constrained by our finite brain capacity, which is not uniquely specified among human beings.

A minority of models and theories are successful, and often only for a limited time.  The scientific process, like biological evolution, is selective.  In this view, the understanding of Nature by human beings looks no longer as much miraculous as Albert Einstein thought\footnote{``The eternal mystery of the world is its comprehensibility\dots The fact that it is comprehensible is a miracle.'' Albert Einstein}. All theories are derived by human beings from the search of a formal simplified description of Nature that is both as faithful as possible, but also that is adapted to the brain of at least some other human beings.  Furthermore, accepted theories are repeatedly checked, scrutinized for correctness by the scientific community, so in the end only effective theories survive.

At this stage it is worth to put some attention to mathematics, which is used as the most solid formal language in physical models and theories.  Mathematics is not a frozen field and its view by mathematicians has deeply changed during the 20$^{\mathrm{th}}$ century.  Since about the Greek school of mathematics around $-300$ BC, and particularly since Euclid, mathematics and logic have been based on a limited set of axioms from which all the theorems were expected to follow with rigorous demonstrations.  But during the 20$^{\mathrm{th}}$ century, famous mathematicians like Kurt G\"odel,
Alan Turing, 
and Gregory Chaitin 
showed however that such an axiomatic mathematics would always be incomplete (see \cite{chaitin} for an introduction to these problems and for references).  Not by a little part, since actually most theorems (``mathematical facts'') escape demonstration when starting from a limited set of axioms.  Since physical theories try to build an isomorphism between Nature and a subset of mathematics, it follows that no ``theory of everything'' (TOE) can summarize with provable derived theorems the implicit complexity of its content.  Similar ideas have been expressed by Stephen Hawking%
\footnote{\texttt{http://www.damtp.cam.ac.uk/strings02/dirac/hawking/}}.  
Therefore even if a theory would be exactly true, most of the consequences of it would escape an axiomatic description.  Only left is the possibility to \textit{compute} by brute force methods the consequences of theories. This is why the exponentially growing capacity of computers plays an increasingly important role in science, because computers allow to explore better and better the vast domain of theories that were out of reach by older methods using mainly theorems (like analytical methods).  The best illustration of the power of the computer approach vs.\ the traditional analytic approach is the very notion of deterministic chaos that arose from computer models in the 1960's \cite{henon,lorenz}.

\section{Chaos in galaxies}
Let us now apply these above general ideas to galaxies.  These particular structures in the Universe offer an excellent case illustrating the scientific method at work applied to complex phenomena where chaos takes an important part.  Galaxies indeed contain a large variety of physical problems at widely different scales, most of them chaotic at some level, which have been more or less successfully been described by distinct models.  No galaxy model may claim to describe fully and exactly the galaxies, but the rich variety of models provide complementary descriptions of galaxies that overall improve our understanding of these objects.  With these models, all deliberately simplified for making a part of galaxies understandable, we can much better think about galaxies than without.

Let us mention some types of chaos in galaxies existing at very different levels.

\subsection{Chaos linked to Newtonian dynamics}  
Before and even after the discovery of interstellar matter and dark matter ($> 1930$), galaxies have been described by purely stellar dynamical models of point masses following classical mechanics, the N-body model.  Such a simplified description already demands substantial efforts to extract useful information helping us to understand how galaxies work. For a long time the mean field approximations, models of the N-body model, were studied, with further time-independence and spherical or axisymmetric symmetry assumptions:
\begin{enumerate}
 \item \textit{Orbit description}. Assuming a fixed, smooth mean field potential allows to decouple the motion of point masses.  At this stage one considers all the possible individual orbits in the galactic potential.  As soon as the potential departs from a spherical shape, or differs from some particular functional form like the St\"ackel potential family (for example due to the presence of a rotating bar), orbital chaos appears and occupies large regions of phase space associated to resonances. The neglected granularity of the mass become then relevant on chaotic orbits,  the relaxation process estimated in integrable potentials is much faster \cite{pfenniger86}.

Over several decades George Contopoulos and collaborators (e.g., \cite{contopoulos80,contopoulos_patsis,patsis,voglis99,voglis07}) have explored  by numerical means the orbital complexity in the phase space of galactic potentials, illustrating well how numerical ``brute force'' orbit integration did allow to make understandable a part of the complexity of the mean field model of galaxies.  In contrast, traditional analytical approaches have been much less efficient in extracting the information implicitly contained in the classical mechanics models of galaxies. 

\item \textit{Phase space fluid description}.
Another approach is to approximate by infinity the large but finite number of stars or weakly interacting dark matter particles in a galaxy.  In addition, these collisionless particles are supposed to be smoothly distributed in phase space, and make a continuous and differentiable flow of matter in the 6-dimensional phase space of space and velocity coordinates.  The differentiability of such a collisionless flow is a strong hypothesis that is by far not justified, neither by the observations of the stars in the solar neighbourhood, where stellar streams abound, nor by the usual arguments used in collisional flows in which smoothness is expected to arise through the microscopic chaos resulting from the frequent particle collisions (the relaxation time is short), erasing quickly irregularities and decorrelating particles.  

Further, one should also keep in mind that the number of stars in galaxies ($<$$10^{12}$) is actually not very large for a smooth fluid description in a  6-dimensional phase space, and lies also in part in massive star clusters.  If one would represent phase space with cells each containing, say, at least 100 particles in order to have reasonably smooth average quantities between contiguous cells, one would obtain a number of bins per coordinate of only $({10^{12}/10^2})^{1/6} \sim 50$.  Dark matter particles are more numerous but are not expected to contribute much density in the optical part of galaxies like the Milky Way, therefore the graininess of the mass distribution in galaxies is certainly already a difference with the smooth model that must be taken into account.

\item \textit{N-body description}.
The full N-body model of a galaxy is a much more faithful model than the two previous ones, at least when one wants to describe a galaxy with about $10^{12}$ point masses.    In a not too distant future, 5--10 years, it is likely that it will be possible to integrate such as number of particles with the forces calculated by tree approximation techniques.  Therefore the N-body model has certainly a bright future and will make the collisionless smooth flow model less relevant in galactic dynamics, and even less in star cluster dynamics.  However the N-body model is also limited.\\
\textbf{a)} The Miller's exponential instability \cite{miller64} of individual particle trajectories and of the whole N-body system constrains to interpret N-body simulations in a statistical way.  An ensemble of close but otherwise uncorrelated initial conditions produces an ensemble of round-off errors dependent evolutions, which may contain statistical useful information about typical representative evolutions.\\
\textbf{b)} Another important limitation of the N-body approach when the bodies are supposed to represent real stars is that one often neglects the internal evolution of stars.  At formation, most of the mass transformed into stars consists of small mass stars that eject a large fraction of their mass in the red giant phase several Gyr later.  This ejected gas from planetary nebulae mixes mass, orbital momentum and kinetic energy in the interstellar gas, a highly energy dissipative process.  Galactic models that ignore this dissipative aspect obviously miss an important part of reality over long time scales, especially in 5--12 Gyr old systems like elliptical galaxies that have been considered for a long time as prototypical systems for using pure dissipationless N-body dynamics.  In fact ellipticals, like spiral galaxies but for different physical reasons, should be seen as substantially dissipative systems when described over several Gyr.
\end{enumerate}

\subsection{The complex physics of baryons}
Besides stars, galaxies contain gas in sometimes large amount, even exceeding the stellar content.  Some very gas rich galaxies like Blue Compact Dwarfs galaxies are misnamed: they are called dwarf only because the visible stellar mass is tiny, but when the gas content revealed by HI emission is considered, they appear just as massive as normal galaxies, except for the fact that gas has not yet been turned into stars. 
So for galaxies even with lesser amount of gas, stellar dynamical models are more or less rapidly invalidated by the rest of the physics that baryons can be subject to.

The interstellar gas physics is very complex and far from being under control.  Typical interstellar gas is multiphased, has supersonic  turbulence, and density and temperature contrasts covering several orders of magnitudes, a very chaotic state that defies description with the present physics tools.  For example, thermodynamics supposes for its use that a local thermal equilibrium can be established, provided a local mechanical equilibrium has been reached. But supersonic turbulence means precisely that strong pressure gradients are ubiquitous, out of mechanical equilibrium regions frequent.  Despite such incoherences, thermodynamical quantities like temperature are used in models and observations due to a lack of better theoretical tools about supersonic compressible turbulence.

What is apparent from simple order of magnitude estimates is that dissipative effects and the exchange of energies between the stars and the gas is not negligible for the whole galaxy equilibrium over Gyr timescales. For example the power radiated by the stars at their different stages of evolution, known to be partly recycled by the dust in the infrared, or known to feed a part of the turbulence in the interstellar medium, is comparable to the power necessary to change the whole galaxy shape against its own gravity \cite{pfenniger91}.  Therefore, the galaxy global parameters and shapes can be expected to depend also on its internal dissipative micro-physics, and not only on the initial conditions at earlier epochs, or external effects like accretion.

\subsection{The dynamics of non-baryonic matter}
Solid cosmological and particle physics arguments exist for the existence of large amount of non-baryonic matter.  1\,s after the Big Bang a number of neutrinos comparable to the photon number must have been  produced mostly from electron-positron annihilations.  The involved physics is the well known, far from exotic MeV nuclear physics.  The discovery of neutrino oscillations between the e-, $\mu$- and $\tau$-neutrinos was a proof of their positive mass, and solved the 40 year old solar neutrino deficit.  With the present constraints about the neutrino mass ($\sim 0.01 -0.1$\,eV), this average leptonic density turns out to be comparable to the average identified baryon density \cite{pfenniger06}.   The neutrino case shows that a particle predicted by Pauli in the 30's for resolving an apparent violation of energy conservation during the $\beta$ decay demanded huge effort to arrive to the present solid conclusion that indeed much matter is in non-baryonic form.  The gained knowledge about neutrinos suddenly doubles the amount of identified matter, which is this time leptonic.

With similar arguments, many other particle candidates (axions, neutralinos, super-symmetric particles, \dots) have been proposed, often with strong theoretical motivations based on symmetries and conservation laws. For example, axions are invoked for explaining the
zero neutron electric dipole moment, an empirical fact that escapes predictions of the Standard Model.  Therefore, in view of all the
oddities remaining to be explained in elementary particle physics, it is natural to expect a rich variety of different dark matter components that remain to be identified.

The consequence for galaxies is that each kind of matter can imprints different effects during structure and galaxy formations.   Some phases of structure formation, like during the formation of Zel'dovich's pancakes, are highly sensitive to the neglected physics.  Yet a high fraction of matter is expected to participate at least once, even over a brief time interval, to a sheet-like singularity where the outcome of such highly non-linear singularities is known to be very sensitive to the exact physics of the participating matter, so also from non-baryonic particles.  The often adopted collisionless property of cold dark matter is just an assumption that may be acceptable in present day galaxy models, but may turn out to be invalid during perturbation sensitive events like pancake or filamentary collapses.

\section{Cases of Sensitive Dependence in Galaxy Models}
In the following we will concentrate on illustrative cases of sensitivity to perturbations of galactic models, where slight changes in the model may
turn out to lead to radically different conclusions.  We will discuss the perfectly smooth fluid phase space description of the star ensemble and of collisionless matter used in the collisionless Boltzmann equation, where the collisionless limit may turn out to lead to severe approximations. The discrete point mass models used in N-body simulations is also limited by its strongly chaotic character. Finally the role of quantum mechanics at extra-galactic scales related with cosmological neutrinos and possibly other relic dark matter particles will be argued to be not so negligible as
usually assumed.

\subsection{Collisionless Chaos}
The main equation of collisionless galactic dynamics is the collisionless Boltzmann equation.  Suppose that we describe the mass density at the instant $t$ in space $x \in \mathbb{R}^3$ and velocity space $v \in \mathbb{R}^3$, i.e., $\{ x,v\} \in \mathbb{R}^6$ by a density distribution $f(t,x,v) \in \mathbb{R}$.  The projection of $f$ onto the $x$-space
\begin{equation}
 \rho(x,t) = \int \! d^3v \, f(x,v,t),
\end{equation}  
provides the usual mass density $\rho(x,t)$.  This projection is well defined even when $f$ is not differentiable. Poisson's equation gives us a constraint on the gravitational acceleration $g$ induced by the mass density $\rho$,
\begin{equation}
 \nabla g(x,t) = -4 \pi G \rho(x,t)  = -4 \pi G \int d^3 \, f(x,v,t) \ .
\end{equation}  
Finally, the collisionless Boltzmann equation 
\begin{equation}
 \partial_t f + v \cdot \partial_x f + g \cdot \partial_v f = 0, 
\end{equation}  
tells us that the mass flow in $\mathbb{R}^6$ is conserved, the characteristics curves of this equation are the trajectories of particles in the acceleration field $g$. The above three equations forms a system of non­linear integro-differential equations, similar to Euler's equation in $\mathbb{R}^3$ for incompressible fluids.

Many efforts have been dedicated by mathematicians to understand the Euler or Navier-Stokes equations with more rigorous tools that commonly used in physics.  This has been useful to understand much more general facts about non-linear partial differential equations, such as the limits of their applicability. 

The simplest case taken from the Navier-Stokes equation but still preserving its non-linear character is the 1-dimensional Burger's equation,
\begin{equation}
 \partial_t u + u \partial_x u = \mu \partial^2_x u \ ,
\end{equation}  
which describes a constant density advection flow at velocity $u(x,t)$ along the direction $x$, with an optional viscosity term proportional to a parameter $\mu$ on the right-hand side.  When $\mu=0$ the system corresponds to the energy conserving Euler equation.  For simple initial conditions, say $u(x,0) = -\sin(x)$, the solution becomes multi-valued after a finite time when $\mu=0$, and develops a shock, a discontinuity when $\mu>0$.  This shows a prototypical behaviour of non-linear partial differential equations: they tend to break the initial assumptions about the solution after a finite time by violating the assumption that the solution remains continuous and differentiable everywhere.  We have here an example of sensitive dependence on the functional form of the flow model.
If viscosity is zero the flow develops multi-valued velocities, while if viscosity is small but positive the flow remains single-valued but becomes discontinuous, and most of the energy is dissipated in the shocks.

An old but relevant result by Onsager \cite{onsager} \footnote{An extensive review about Onsager's work on turbulence is given in \cite{eyink}.} about 3-dimensional turbulence is that energy dissipation in Navier-Stokes fluids does not vanish to zero when viscosity tends toward zero.  As viscosity tends toward zero, Navier-Stokes'  \textit{equation} tends well toward  Euler's equation, but the \textit{solutions} don't.  The most astonishing fact verified in experiments of developed turbulence is that the energy dissipation tends toward a positive constant independently of the value of viscosity. As in Burger's equation entropy increasing discontinuous solutions (so-called ``weak'' solutions) are physically the relevant ones.

Therefore it may be physically misleading to consider Euler's equation and its energy conservation law, for application on systems where viscosity is small, because even a small viscosity term becomes essential when the flow becomes discontinuous, highly turbulent, which is precisely the rule in low viscosity fluids.

There is no ground to believe that the growth of discontinuities is
restricted to non-linear hydrodynamics.  On the contrary, the growth
of shocks and discontinuous solutions occur  frequently in other non-linear partial differential equations.   One should expect shocks in more complex, higher dimensional non-linear systems like the Poisson-Boltzmann system, in which  the collisional or diffusive term is small, but is never exactly zero.  In cases of strong phase space ``turbulence'' during collapses and violent events we can expect that the small residual collisionality has its diffusive effects strongly amplified.  High phase space density gradients and multi-streams are susceptible to develop fast from the actual particle noise. 

Also we should remind that differentiability in usual fluids is often a valid assumption because the microscopic molecular chaos does erase the growth of correlations faster than the flow develops gradients of density or velocity.  Except for very particular cases, laminar flows do require some positive viscosity, low Reynold's number, to stay laminar.  Otherwise low viscosity fluids become spontaneously turbulent, i.e., develop discontinuous, singular flows. 

Since precisely collisionless flows in galactic dynamics lack of a strong microscopic collisional chaos that would justify the usual smoothness assumption of distribution functions, we should rather expect irregular, non-smooth distributions (``weak'' solutions) as a rule in galaxies.  Actually, the local distribution function of stars in the solar neighbourhood is highly structured with several star streams, and is far from resembling a Maxwellian distribution.  What we see is at best a  partly relaxed distribution.   

To explain the smoothness in the galaxy distribution functions, the violent relaxation concept \cite{lynden-bell} has been proposed.  It was initially attributed to the time-dependence of the gravitational potential, but today we view it rather as resulting from the highly sensitive to perturbations, chaotic stages of galaxy evolution where microscopic perturbations become fast macroscopically relevant.  Contrary to a still popular opinion, this is not directly the time-dependence of the potential that leads to an enhanced relaxation, but the highly chaotic, sensitive stage of the system.  Counter-examples demonstrating that time-dependence does not necessarily relax a collisionless flow are analytical time-dependent periodic solutions of the Poisson-Boltzmann system (e.g., \cite{sridhar}).

\subsection{N-body Chaos}
Miller (1964) \cite{miller64} discovered numerically the exponential divergence of particular gravitational N-body systems, and noticed that the divergence of close systems occurs not because of transient close 2-body encounters, but constantly by the $N(N-1)$ particle interactions.  Gurzadyan \& Savvidy (1986) \cite{gurzadian} showed also with Riemannian geometry that the N-body problem is indeed generally chaotic in simple particle distributions.  However these studies should not be seen as definitive, it is indeed not difficult to invent particular counter-examples of as weakly unstable as wished configurations of N-body systems, such as widely separated pairs of pairs etc. of binaries.  By natural selection we do observe in the sky the least unstable multiple star systems, often arranged hierarchically. The solar system is also an example where the Liapunov time is much longer than its dynamical  time..

An interesting problem is to specify the relationship between Miller's type chaos seen at the microscopic level, and the global stability of a stellar system.   In usual gases, the molecular very rapid chaos is the key property that guarantees that the system seen at macroscopic scales can be modelled with the quasi-deterministic rules of thermodynamics.  To be effective thermodynamics requires a fast relaxation of molecules, in other words, a strong molecular chaos.  Is it similar in gravitational systems?
From numerical experiments, systems like hot spherical models of many equal mass stars are examples where indeed a kind of statistical robust state appears to be reached over time-scales longer than the crossing time.   For these systems Miller's microscopic chaos could be actually favouring a global statistical quasi-equilibrium.

Other gravitational systems, such as disks, or systems with strongly anisotropic velocity dispersions, can present macroscopic instabilities leading to evolution.   It is presently unclear whether Miller's microscopic chaos is related in any way with large scale instabilities.

\subsection{Sensitivity to Quantum Physics} 
The relict cosmological neutrinos are fermions and have been produced $\sim 1-2$\,s after the Big Bang from electron-positron annihilations at a redshift of $z \approx 10^{10}$.  At this epoch all the particles were strongly relativistic and close to thermal equilibrium.  This means that the distribution of neutrinos had to be very close to a Fermi-Dirac distribution with a negligible chemical potential.  In other words, the natural creation state of such relativistic particles is always  semi-degenerate \cite{weinberg}, which means that quantum mechanics plays a significant role from the start.  The  phase space density of neutrinos is therefore sufficiently high for quantum effects to be important at the macroscopic level.  As the Universe expands the neutrino phase space density is little modified, since  neutrinos have only two weak possibilities of interaction with matter, the weak force and the gravitational force, and their number cannot decrease through decay since they are supposed to be stable.

For a rest mass range such as $0.01-0.1$\,eV  at the present time the relict neutrinos should be non-relativistic and much colder than the often quoted temperature of 1.9\,K, which would apply if they were massless or still relativistic \cite{pfenniger06}.   Their speed is estimated low enough ($\sim 1000 \,\rm km\, s^{-1}$) to be trapped at least in galaxy clusters.  So if neutrinos are massive and participate to structure formation, clearly they are able to perturb the outcome of pancake collapses since their total mass is comparable to the one of baryons, and they are still today semi-degenerate fermions, so their physics differs from classical ballistic particles.  Structure formation is a highly chaotic process where most of the matter goes through Zel'dovich's pancakes, so one can expect that any perturbation to a purely classical model of structure formation, like  when including the fermionic properties of neutrinos, may significantly change the results.

But here we have the non-trivial challenge to merge two widely different descriptions of Nature.  Sometimes quantum physics is said to be ``holistic'' because it is a non-local theory. A ``particle'' does not necessarily represent a localized mass, but can be a plane-wave.  Pauli's principle does not require to localize particles, and the wave-function in Schr\"odinger equation may extend over all space.  In contrast, the notion of localized point mass is central in classical mechanics. 

In the recent years these questions have considerably advanced.  The decoherence theory \cite{zeh70,zeh05,zurek02,zurek03} allows to specify when a finite system, supposed to be isolated, is well represented by quantum physics, and when classical physics is the best description: 
when the neglected but always existing perturbation rate of the outer world to \textit{a given model} stays indeed negligible, quantum coherence build-up wins and quantum mechanics stays a faithful representation of the system. But when the outer world perturbation rate is faster and destroys  quantum coherence, classical physics emerges as a good description of the system.  A satisfactory aspect of the decoherence theory is that a conscious observer is no longer required.  A measurement corresponds to coherence destruction by external perturbations to the system.  Such considerations are central for building quantum computers, where coherence preservation is crucial, and clearly unrelated with the presence of a conscious observer. 

So in order to describe a particle as a localized mass distribution with  classical mechanics, as implicit in all cosmological simulations up to now, one needs to check that decoherence is effective. Otherwise particles can not be localized and must be described with quantum physics as non-local ensembles. 

But since neutrinos have extremely low probability to interact with other  particles by the weak interaction, the only remaining possibility is gravitational interaction.  As long as the Universe is homogeneous, gravitational interaction cancels and is negligible.  Only when structures form ($z \ll 100$) matter inhomogeneities perturb neutrinos by gravitation. 

The entanglement time-scale $\tau_E$ of identical neutrinos is estimated  by a classical collision time where the cross section diameter is given by the de Broglie wavelength $\lambda_{\rm dB} = h/m_\nu v $ (see \cite{pfenniger06}),
\begin{equation}
  \tau_E \approx (n_\nu \lambda_{\rm dB}^2 v_\nu)^{-1} 
         \approx 1.3 \cdot 10^{-8} (1+z)^{-2} 
                 \left( \frac{n_\nu}{56} \right)^{-1}  
                 \left( \frac{m_\nu}{1 \rm eV} \right)^{-1} \rm s, 
\end{equation} 
where $n_\nu$ is the neutrino density, $m_\nu$ the neutrino mass, and 
$z$ the redshift.  After a multiple of $\tau_E$ large numbers of neutrinos are entangled just because they are identical fermions, not because of their weak interaction. 

The decoherence theory \cite{zeh70,zeh05,zurek02,zurek03} gives an estimate of the decoherence time $\tau_D$ if we know the characteristic relaxation time $\tau_R$ over which the external world perturbs the system of size $\Delta x$: 
\begin{equation}
        \frac{\tau_D}{\tau_R} = 
                   \left(\frac{\lambda_{\rm dB}}{\Delta x} \right)^2.
\end{equation}
For ensembles of neutrinos above a given size $\Delta x$ the decoherence time $\tau_D$ will be shorter than the entanglement time $\tau_E$, and above such a scale the neutrino ensemble may be considered as localized and included in a classical description, like hydrodynamics. 

In \cite{pfenniger06} we estimated that at the present epoch the fastest relaxing mass condensations for neutrinos are the galaxies, not the stars or the galaxy clusters.  With this estimate of the ``relaxation'' time $\tau_R$ produced by the external world, we found that when considering scales much larger than $\sim 10^{13}$\,cm, the ensemble of at least $10^{40}$ neutrinos is perturbed at a rate fast enough by the galaxy gravitational interactions to be considered as a classical fluid with a Fermi-Dirac equation of state.  Thus a simple model of cosmological neutrinos above solar system scales is to describe them as a Fermi-Dirac fluid, like what is done in stellar models of white dwarfs and neutron stars.  This is a very different physics than the collisionless classical mechanics representation that up to now has been used in cosmological simulations including neutrinos, but closer to the adhesion model sometimes used in pancake models.   Of course the consequences for structure formation models may be drastic because fluids tend to develop shocks, contrary to collisionless flows, and shocks imply entropy production and dissipation. 

What has been said about neutrinos may be applicable to other dark matter particles like axions or neutralinos, with possible complications for bosons.  Much depends on their effective rest mass, number density, and if they are sufficiently non-relativistic to participate to structure formation.

\section{Conclusions}

The sensitivity to perturbations is a characteristics of chaos. Various aspects of galaxies can be represented by different models that may contain sensitive parts subject to limitations on the scope of applicability. 
 
Here are a few points to consider for future galaxy models: 
\begin{itemize}
 \item The assumed smoothness of distribution functions for collisionless systems is not grounded on theoretical arguments or observational evidences.  The effects of irregular, non-smooth distributions on the dynamics of a model have been little investigated.  The results of incompressible fluid turbulence could be useful to extend to collisionless phase space flows.
 \item Therefore the collisionless Boltzmann differential equation is a decreasingly  attractive model in galactic dynamics in regard of the fast growing capabilities of the much less constrained N-body techniques.
 \item The dissipative baryonic physics in galaxies coming from the gas but also from the stellar evolution is important over Gyr timescale, not only for spirals, but also for ellipticals due to the stellar important mass loss in the red giant phase.
 \item Quantum physics in the semi-degenerate sea of cosmological neutrinos produces a kind of collisional relaxation due to decoherence and the fermionic nature of such particles.  These neutrinos and possibly other dark matter particles should not be modelled as collisionless, but as a collisional fluid with the proper quantum statistics equation of state. This point is crucial during pancake shocks.
\end{itemize}

\subsection*{Acknowledgements}
I am grateful to the organisers of this stimulating interdisciplinary conference for the invitation.  This work has been supported by the Swiss National Science Foundation.

\end{document}